\documentclass{PoS}

   \usepackage{color}

\title{Spectral Variability Signatures of Relativistic Shocks in Blazars}

\ShortTitle{Shocks in Relativistic Jets}

\author{\speaker{Markus B\"ottcher}\\
        Centre for Space Research, North-West University, Potchefstroom, 2520, South Africa\\
        E-mail: \email{Markus.Bottcher@nwu.ac.za}}

\author{Matthew G. Baring\\
        Department of Physics and Astronomy, Rice University, Houston, TX 77005-1892, USA\\
        E-mail: \email{baring@rice.edu}}       
        
\abstract{Mildly relativistic, oblique shocks are frequently invoked as possible 
sites of relativistic particle acceleration and production of strongly variable, polarized 
multi-wavelength emission from relativistic jet sources such as blazars, via diffusive shock 
acceleration (DSA). In recent work, we had self-consistently coupled DSA and radiation transfer 
simulations in blazar jets.  These one-zone models determined that the observed spectral energy 
distributions (SEDs) of blazars strongly constrain the nature of the hydromagnetic turbulence 
responsible for pitch-angle scattering. In this paper, we expand our previous work by including 
full time dependence and treating two emission zones, one being the site of acceleration. 
This modeling is applied to a multiwavelength flare of the 
flat spectrum radio quasar 3C~279, fitting snap-shot SEDs and light curves. We predict 
spectral hysteresis patterns in various energy bands as well as cross-band time lags
with optical and GeV $\gamma$-rays as well as radio and X-rays tracing each other closely
with zero time lag, but radio and X-rays lagging behind the optical and $\gamma$-ray 
variability by several hours. }

\FullConference{6th Annual Conference on High Energy Astrophysics in Southern Africa\\
		1 -- 3 August, 2018\\
		Parys, South Africa}

\begin{document}

\def\actionitem#1{\textcolor{blue}{#1}}  

\section{Introduction}

Relativistic, oblique shocks have long been considered as one of the leading contenders for the 
sites of relativistic particle acceleration in relativistic jet sources, such as blazars and 
gamma-ray bursts, resulting in the observed rapidly variable, often highly polarized multi-wavelength (MW)
emission. The dominant particle acceleration mechanism at such shocks is referred to as diffusive 
shock acceleration (DSA). Particle acceleration results from repeated shock crossings of particles 
gyrating along large-scale ordered magnetic fields. The reversal of particle momenta $p$ along magnetic
field lines is facilitated by diffusive pitch-angle scatterings (PAS). Several theoretical studies of particle 
acceleration at relativistic shocks (e.g., \cite{KH89,Ellison90,ED04,SB12}) have shown that this 
process can result in a wide variety of spectral indices. Such studies of the particle acceleration 
mechanism, however, usually do not consider the resulting radiative signatures in a self-consistent manner. 

On the other hand, models focusing on the multi-zone radiative transfer problem for internal-shock models
of blazars (e.g., \cite{MG85,Spada01,Sokolov04,Mimica04,SM05,Graff08,BD10,JB11,Chen11,Chen12}) do not typically 
address the details of particle acceleration, but assume an ad-hoc injection of purely non-thermal relativistic 
particles, usually with a truncated power-law distribution in energy. In recent work \cite{Baring17}, we 
coupled the Monte Carlo (MC) simulations of DSA of Summerlin \& Baring \cite{SB12} with radiative transfer 
routines of B\"ottcher et al. \cite{Boettcher13}. This provided, for the first time, a consistent description 
of the DSA process  and its radiative signatures in mildly relativistic, oblique shocks in blazar jets. Fits 
to spectral energy distributions (SEDs) of three blazars indicated the need for a strongly energy-dependent 
PAS diffusive mean-free path $\lambda_{\rm pas} \propto p^{\alpha}$, with
$\alpha \sim 2$ --- 3 required, depending on the type of blazar considered. This may be considered as evidence
of hydromagnetic turbulence levels gradually decreasing with increasing distance from the shock 
\cite{Baring17,BB17}. 

In this work, we present an extension of the DSA + radiaton-transfer model of \cite{Baring17}, including
full time variability. We thus make predictions for time-dependent snap-shot SEDs and multi-wavelength light
curves which can be further analyzed to predict multi-wavelength spectral hysteresis patterns and inter-band
time lags. In Section \ref{scheme}, we describe our model setup and the numerical scheme we developed for 
simulating time-dependent DSA and radiation transfer in internal shocks in blazars. The application to a
multi-wavelength flare of 3C~279 is presented in Section \ref{3C279}, yielding fairly good MW spectral fits
and distinctive temporal characteristics.

\section{\label{scheme}Setup and Numerical Scheme}

Our time-dependent shock-acceleration and radiation-transfer simulations are based on the premise that
a DSA-type particle acceleration mechanism is at work in the high-energy emission region of a blazar 
jet at all times. A quiescent state is established through a balance between time-independent DSA in 
a small acceleration zone and radiative cooling and escape of particles in a larger radiation zone
of length $\ell_{\rm rad}$, which is identified with the high-energy emission region, as described in 
detail in \cite{Baring17}. Variability arises from the passage of a mildly relativistic shock through 
the density and magnetic field structures in the high-energy emission region,
nominally on an observed time scale $\Delta t_{\rm obs} = (\ell_{\rm rad} / v_s)
\, (1 + z) / \delta$.  Here $v_s$ is the shock velocity in the co-moving frame of the jet material,
$z$ is the cosmological redshift of the source, and $\delta$ is the Doppler factor arising from
the bulk motion of the jet material with respect to the observer. 

The thermal + non-thermal particle distributions resulting from DSA have been evaluated using the 
Monte Carlo (MC) code of \cite{SB12}. In the DSA scenario, the Fermi-I acceleration process that 
includes episodes of shock drift energization is facilitated by stochastic PAS
of charges spiraling along magnetic field lines. PAS is parameterized through the 
corresponding mean-free path $\lambda_{\rm pas}$ as an energy-dependent multiple $\eta (p)$ of the 
particle's gyro radius, $r_g = p c / (q B)$, where $p$ is the particle's momentum, such that 
$\lambda_{\rm pas} = \eta (p) \, r_g$. The energy dependence of the mean-free-path parameter 
$\eta$ is defined as a power-law in the particle's momentum, $\eta (p) = \eta_1 \, p^{\alpha - 1}$, 
so that $\lambda_{\rm pas} \propto p^{\alpha}$ and $\eta_1$ describes the mean free path 
in the non-relativistic limit, $\gamma \to 1$.  

The MC simulations of \cite{SB12} illustrate that DSA leads to a non-thermal power-law tail
of relativistic particles which have been accelerated out of the remaining thermal pool.  A high-energy 
cut-off ($\gamma_{\rm max}$) of the non-thermal particle spectra results from the balance of the 
acceleration time scale $t_{\rm acc} (\gamma_{\rm max}) = \eta (\gamma_{\rm max}) \, t_{\rm gyr} (\gamma_{\rm max})$ 
with the radiative energy loss time scale. If synchrotron losses dominate, $\gamma_{\rm max} \propto B^{-1/2}$. 
This will lead to a synchrotron peak energy $E_{\rm sy} \sim 240 \, \delta \, \eta^{-1} (\gamma_{\rm max})$~MeV. 
Notably, this synchrotron peak energy is independent of the magnetic field $B$, as $E_{\rm sy} \propto 
B \, \gamma_{\rm max}^2$. Blazars typically show synchrotron peaks in the IR to soft X-rays. In order to
reproduce these, the pitch angle scattering mean-free-path parameter $\eta (\gamma_{\rm max})$ has to 
assume values of $\sim 10^4$ -- $10^8$. However, 
\cite{SB12} have shown that $\eta_1$ must be significantly smaller than this value in order to obtain 
efficient injection of particles out of the thermal pool into the non-thermal acceleration process. 
From these arguments we can infer that $\eta(\gamma)$ must be strongly energy dependent \cite{Baring17}.  

\begin{figure}[ht]
\centering
\centerline{\includegraphics[width=6.5cm]{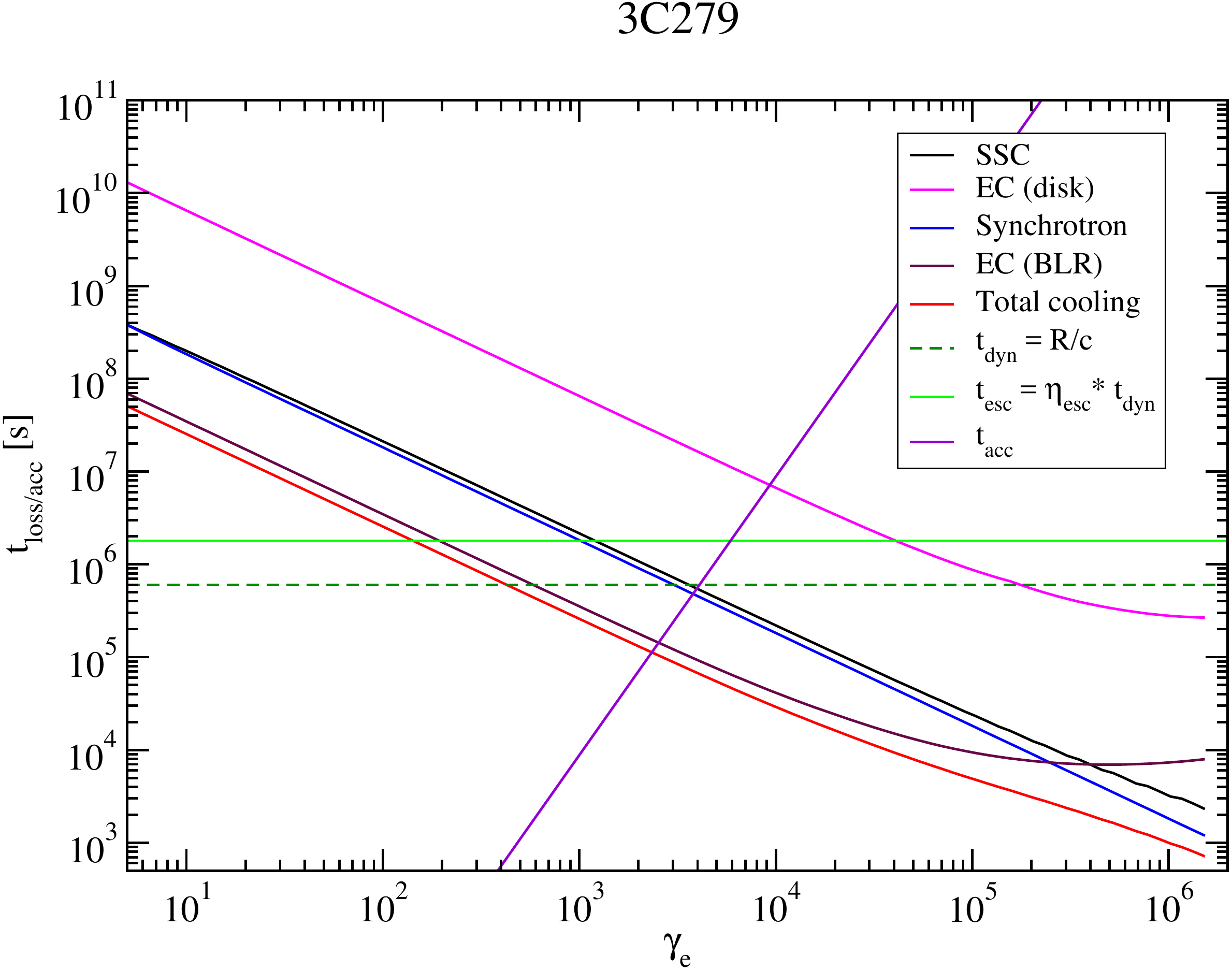} \hskip 10pt \includegraphics[width=6.5cm]{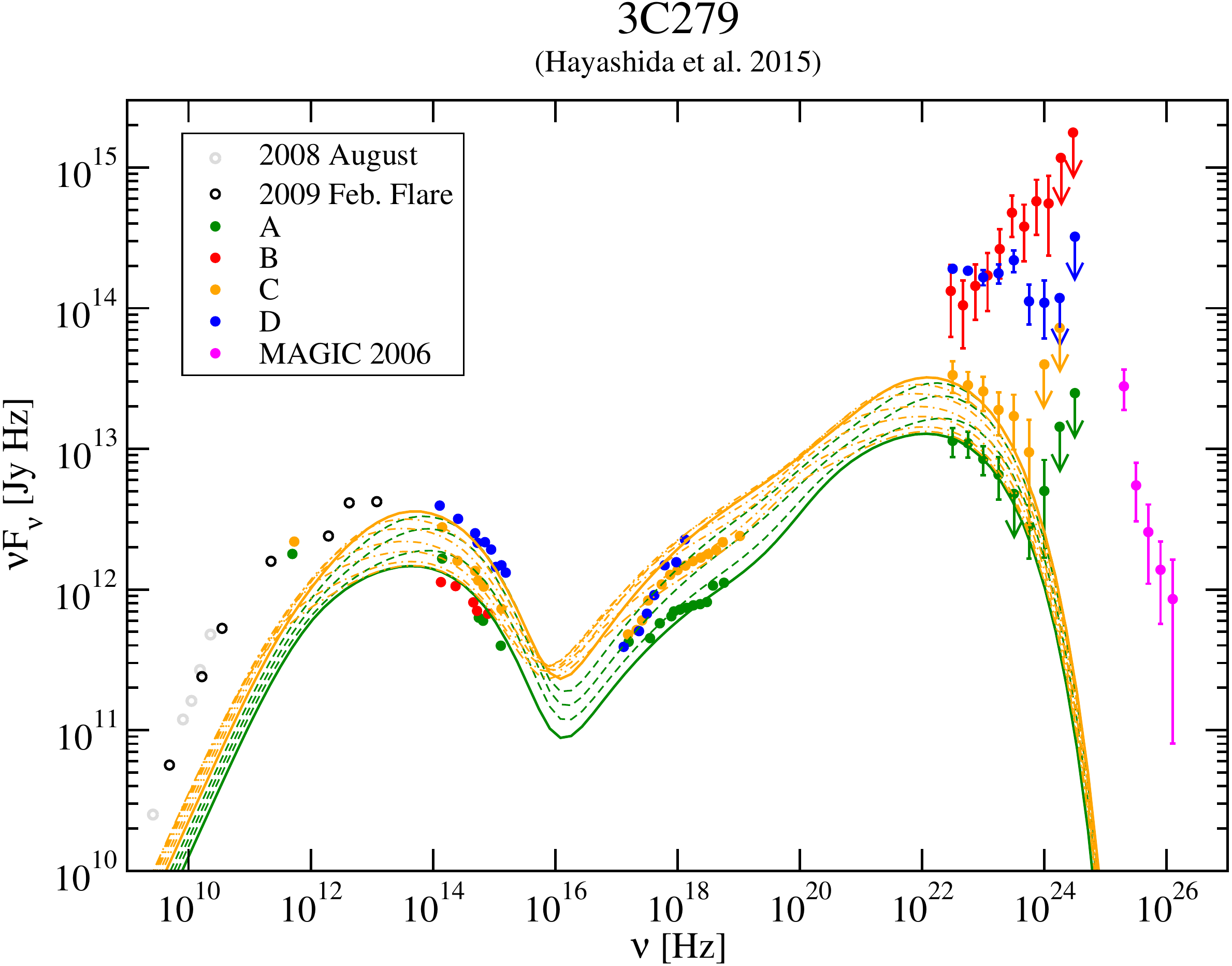}}
\caption{{\it Left panel}: 
Relevant acceleration (purple), radiative cooling (black, pink, blue, brown, red --- see legend),
dynamical (green dashed) and escape (light green) time scales for electrons in the simulated quiescent-state 
equilibrium configuration for 3C~279 (see Section \ref{3C279}). 
{\it Right panel}: 
Snap-shot SEDs of 3C279 during 2013 -- 2014. Data are from \cite{Hayashida15}. Green curves illustrate
the spectral evolution during the rising part of the simulated Flare C; yellow curves show the evolution during
the decaying part. See text for details. }
\label{timescales_SEDs}
\end{figure}

The DSA-generated thermal + non-thermal electron spectra serve as a particle injection term into
simulations of subsequent radiative cooling of the electrons. As relevant radiative mechanisms, 
synchrotron radiation in a tangled magnetic field, synchrotron self-Compton (SSC) radiation, and 
Compton scattering of external radiation fields (external Compton = EC) on various plausible target 
photon fields are taken into account in our simulations. Particles may also leave the emission region 
on a time scale parameterized as a multiple of the light-crossing time scale of the emission region, 
$t_{\rm esc, e} = \eta_{\rm esc} \, \ell_{\rm rad} / c$. Figure \ref{timescales_SEDs} shows the energy 
dependence of the relevant time scales for the steady state generated to describe the quiescent-state 
multi-wavelength emission of 3C~279 (see Section \ref{3C279}). DSA will be effective 
up to an energy $\gamma_{\rm max}$, where the radiative cooling time scale becomes shorter than the 
acceleration time scale. Figure \ref{timescales_SEDs} shows that for almost all particles at lower energies,
$\gamma < \gamma_{\rm max}$, the acceleration time scale is many orders of magnitude shorter than the
radiative cooling and/or or escape timescales. Thus, numerically, DSA may be well represented as instantaneous
injection of relativistic particles at a (time-dependent) rate $Q_e (\gamma_e, t)$ [cm$^{-3}$ s$^{-1}$], 
which is then followed by evolution on the radiative and escape time scales. 

The evolution of the electron distribution is simulated by numerically solving a Fokker-Planck equation
of the form
\begin{equation}
{\partial n_e (\gamma_e, t) \over \partial t} = - {\partial \over \partial \gamma_e} \Bigl( \dot\gamma_e
\, n_e [\gamma_e, t] \Bigr) - {n_e (\gamma_e, t) \over t_{\rm esc, e}} + Q_e (\gamma_e, t) 
\label{FP}
\end{equation}
using an implicit scheme as described in \cite{BC02}. Here, $\dot\gamma_e$ represents the combined radiative 
energy loss rate of the electrons, and all quantities are in the co-moving frame of the emission region. 
Radiation transfer is being handled by forward evolution of a continuity equation for the photons,
\begin{equation}
{\partial n_{\rm ph} (\epsilon, t) \over \partial t} = {4 \, \pi \, j_{\epsilon} \over \epsilon \, m_e c^2} 
- c \, \kappa_{\epsilon} \, n_{\rm ph} (\epsilon, t) - {n_{\rm ph} (\epsilon, t) \over t_{\rm esc, ph}} \quad ,
\label{radiation}
\end{equation}
where $\j_{\epsilon}$ and $\kappa_{\epsilon}$ are the emissivity and absorption coefficient, respectively,
$\epsilon = h \nu / (m_e c^2)$ is the dimensionless photon energy, and $t_{\rm esc, ph}$ is the photon escape
time scale, $t_{\rm esc, ph} = (4/3) \, \ell_{\rm rad} / c$ for a spherical geometry \cite{Boettcher97}. 
Radiative processes are evaluated using the routines of \cite{Boettcher13}. The observed flux is provided
by the escaping photons, such that
\begin{equation}
\nu F_{\nu}^{\rm obs} (\nu_{\rm obs}, t_{\rm obs}) = {\epsilon^2 \, m_e c^2 \, n_{\rm ph} (\epsilon, t) \,
\delta^4 \, V_{\rm rad} \over 4 \pi \, d_L^2 \, (1 + z) \, t_{\rm esc, ph}} \quad ,
\label{nuFnu}
\end{equation}
where $\epsilon = (1 + z) \epsilon_{\rm obs} / \delta$ and $V_{\rm rad} \approx (4/3) \, \pi \, \ell_{\rm rad}^3$
is the co-moving volume of the emission region, and jet-frame and observer
time intervals are related through $\Delta t_{\rm obs} =
\Delta t \, (1 + z) / \delta$. 

Our code produces snap-shot SEDs and multi-wavelength light curves at pre-specified frequencies. It also
extracts local spectral indices at the same frequencies for each time step, for the purpose
of plotting hardness-intensity diagrams. Correlations between the light curves at different frequencies
and possible inter-band time lags are evaluated using the Discrete Correlation Function analysis \cite{EK88}. 
For each flare simulation, we first let the code run until it reaches a stable equilibrium with the 
quiescent-state parameters. An individual flaring event is then simulated by changing various input 
parameters with a step function in time for the duration $\Delta t = \ell_{\rm rad} / v_s$ . 

\section{\label{3C279}Application to 3C~279}

As an application of our code, we provide a fit to a multi-wavelength flare of the well-known Flat Spectrum Radio
Quasar 3C~279. Hayashida et al. \cite{Hayashida15} identified several individual flaring episodes in 2013 -- 2014.
For the purpose of this study we select Flare C, which showed simultaneous flaring in the optical, X-ray, and $\gamma$-rays
and may therefore well be represented by an increase in the particle injection luminosity, plausibly caused by an internal
shock in the jet of 3C~279. The characteristic time scale of short-term flares of 3C~279 during the 2013 -- 2014 period 
(includng Flare C) is $\Delta t_{\rm obs} \sim 1$~day. With a typical Doppler factor of $\delta = 15$, a redshift of 
$z = 0.536$, and a mildly relativistic shock with $v_s \sim 0.7$~c, this implies a size of the active region of 
$\ell_{\rm rad} \sim 1.8 \times 10^{16}$~cm. We assume a viewing angle of $\theta_{\rm obs} \approx 1/\Gamma$, so 
that $\delta \approx \Gamma = 15$. 

\begin{table}[ht]
\caption{\label{parameters}Parameters for the model fit to 3C~279.}
\smallskip
\begin{center}
\begin{tabular}{cc}
\hline
\noalign{\smallskip}
Parameter & Value \\
\noalign{\smallskip}
\hline
\noalign{\smallskip}
Quiescent electron injection luminosity & $L_{\rm inj, qu} = 1.1 \times 10^{43}$~erg~s$^{-1}$ \\
Flaring electron injection luminosity & $L_{\rm inj, fl} = 5.0 \times 10^{43}$~erg~s$^{-1}$ \\
Emission region size & $\ell_{\rm rad} = 1.8 \times 10^{16}$~cm \\
Jet-frame magnetic field & $B = 0.65$~G \\
Escape time scale parameter & $\eta_{\rm esc} = 3$ \\
Pitch-angle scattering m.f.p. parameter & $\eta_1 = 100$ \\
Pitch-angle scattering m.f.p. scaling index & $\alpha = 3$ \\
Bulk Lorentz factor & $\Gamma = \delta = 15$ \\
Accretion-disk luminosity & $L_d = 6 \times 10^{45}$~erg~s$^{-1}$ \\
Distance of active region from BH & $z_i = 0.1$~pc \\
External radiation field energy density & $u_{\rm ext} = 4 \times 10^{-4}$~erg~cm$^{-3}$ \\
External radiation field BB temperature & $T_{\rm ext} = 300$~K. \\
\end{tabular}
\end{center}
\end{table}

Both the direct accretion-disk radiation field and an isotropic external radiation field are needed
to model the $\gamma$-ray spectrum of 3C~279. The latter is assumed to be dominated by the 
dust-torus radiation field, which is approximated as a thermal blackbody at a temperature of $T_{\rm ext} = 300$~K. 
The most relevant parameters are listed in Table \ref{parameters}. The quiescent state fit is illustrated by the solid
green line in Fig. \ref{timescales_SEDs}. We find that it can be well described with an electron injection spectrum produced
by DSA with a pitch-angle scattering mean free path scaling as $\lambda_{\rm pas} = 100 \, r_g \, p^2$, i.e.,
$\lambda_{\rm pas} \propto p^3$. Based on the competition of acceleration and cooling time scales, as illustrated
in Fig. \ref{timescales_SEDs}, electrons are accelerated up to a maximum energy of $\gamma_{\rm max} = 2.4 \times 10^3$. 

\begin{figure}[ht]
\centering
\includegraphics[width=6cm]{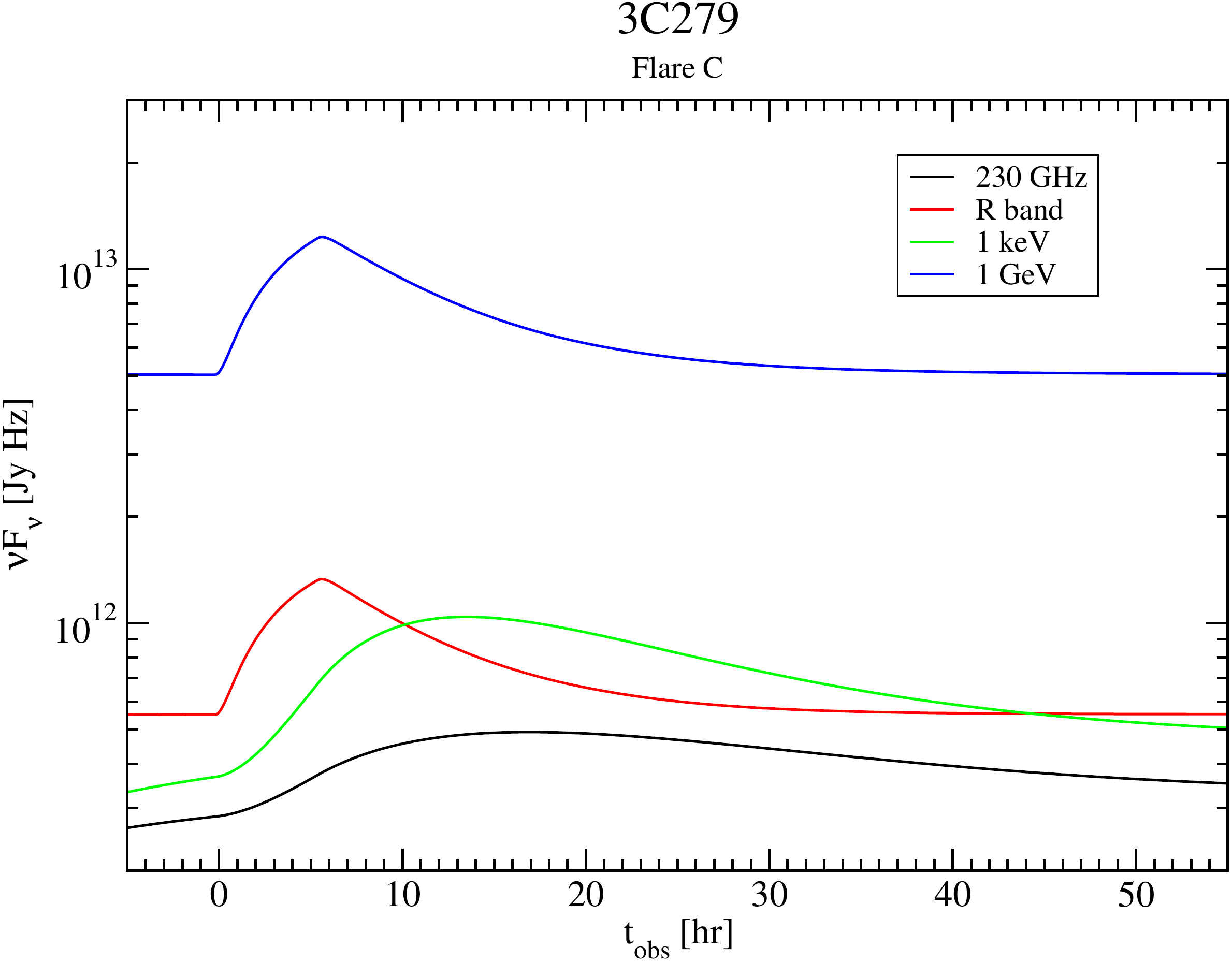}
\quad
\includegraphics[width=6cm]{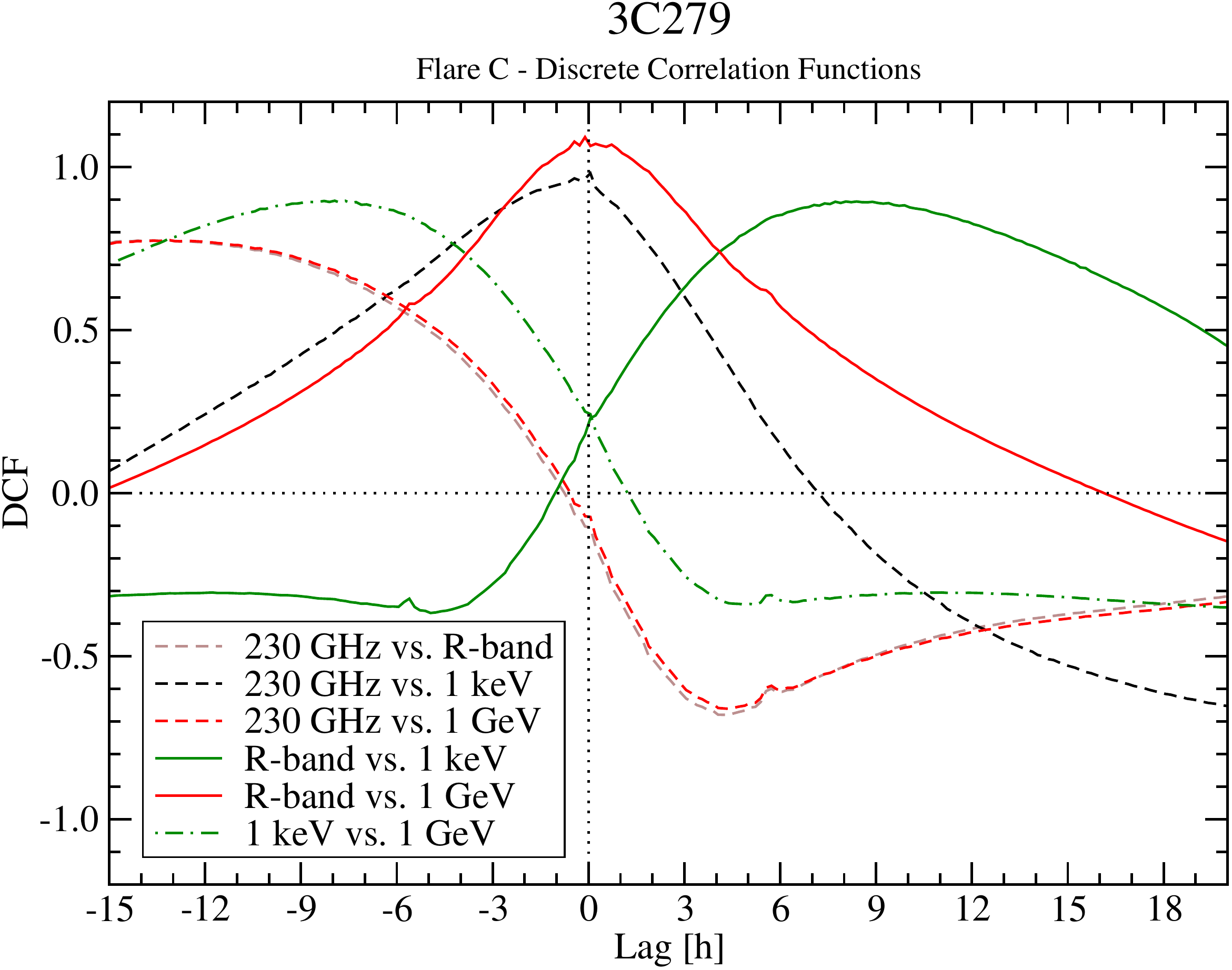}
\caption{{\it Left:} Multi-wavelength light curves extracted from the simulation illustrated in Fig.
\ref{timescales_SEDs}. {\it Right:} Discrete cross correlation functions evaluated from the light curves shown 
in the left panel. }
\label{LC_DCF}
\end{figure}   

For the evolutionary model of flare C, the only parameter changed to produce the flare is the electron injection luminosity, 
corresponding to a larger number of electrons accelerated per unit time, without changing the characteristics of 
the acceleration process. The green curves in Fig. \ref{timescales_SEDs} show SEDs during the rising phase of the flare, the 
solid yellow curve indicates the snap-shot SED during the peak of flare C, while the remaining yellow curves 
illustrate the decaying part of the flare. The resulting light curves in the mm radio, optical, X-ray and GeV
$\gamma$-ray bands are illustrated in the left panel of Fig. \ref{LC_DCF}, while cross-correlations are shown
in the right panel of Fig. \ref{LC_DCF}. 

The model predicts, as expected, that the optical and $\gamma$-ray light curves are closely correlated with 
zero time lag, as those bands are produced by synchrotron and Compton emission from electrons of similar
energies and therefore comparable cooling times. The X-ray emission is expected to lag behind the optical 
and $\gamma$-ray emissions by $\sim 8$~hr,
while the mm radio band is expected to show an even longer delay behind optical and $\gamma$-rays, with 
slightly weaker correlation. Unfortunately, the light curve coverage in existing data (including those for 
3C~279 in \cite{Hayashida15}) is not sufficient for a meaningful comparison of our predictions with data.

\begin{figure}[ht]
\centering
\includegraphics[width=7cm]{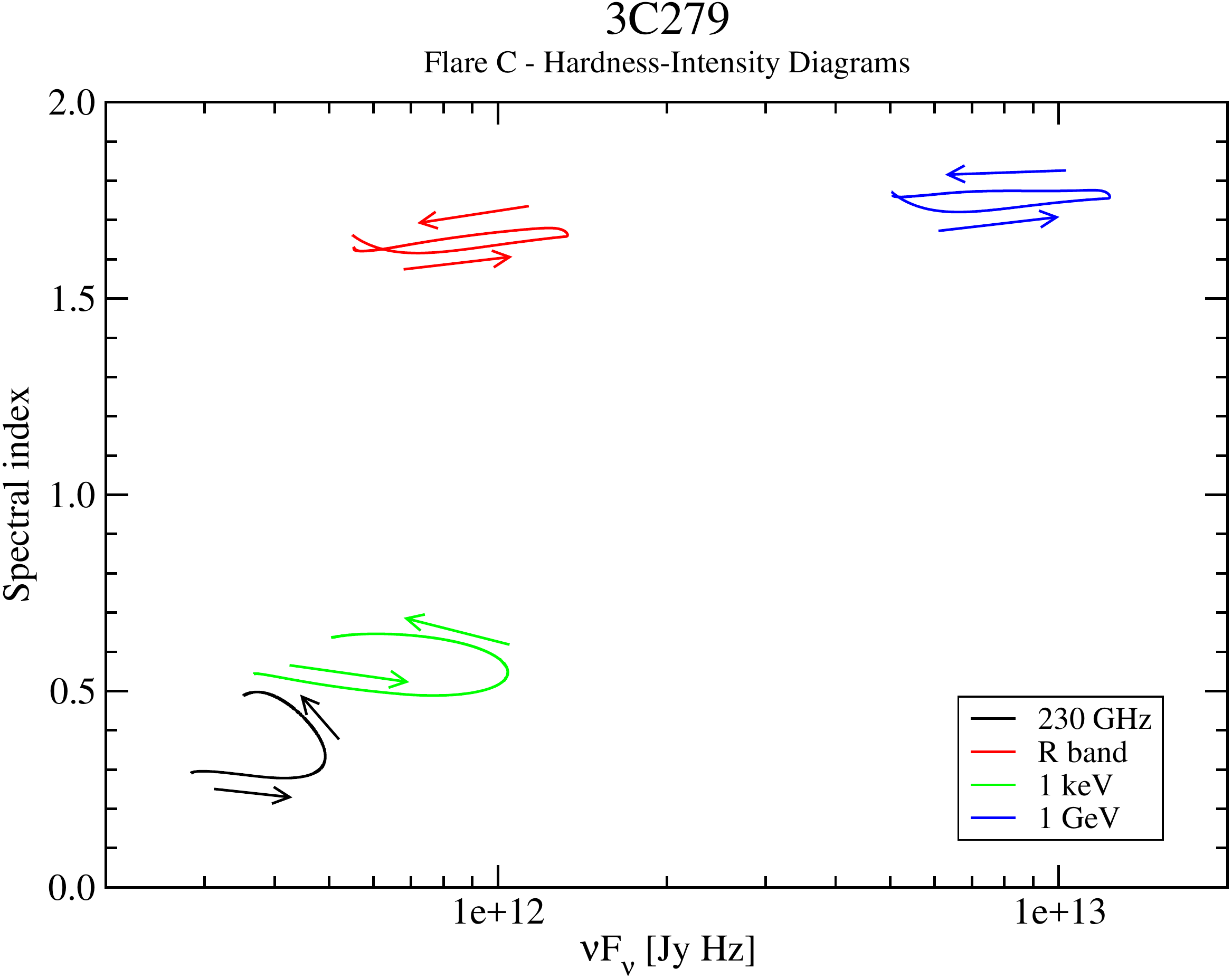}
\caption{Hardness-intensity diagrams extracted from the same flare simulation illustrated in Figs. \ref{timescales_SEDs}
and \ref{LC_DCF}. }
\label{HIDs}
\end{figure}   

Figure \ref{HIDs} shows the hardness-intensity diagrams extracted from our simulatons to flare C. 
While only very weak spectral variability is predicted in the optical and GeV $\gamma$-ray bands,
pronounced counter-clockwise spectral hysteresis (harder rising-flux spectra; softer decaying-flux
spectra) is expected in the mm radio and X-ray bands. Such spectral hysteresis has so far only been
clearly identified in the X-ray spectra of high-frequency peaked BL Lac objects (e.g., \cite{Takahashi96}),
where the X-ray emission is synchrotron-dominated. Observing such features in other wavelength bands would 
enable stringent constraints on the magnetic field in the emission region (see, e.g., \cite{Boettcher03}). 
Such a goal is quite challenging in the GeV band because of limited count statistics in {\it Fermi}-LAT 
observations.

\noindent {\bf Summary}: 
We have presented the first time-dependent coupled DSA + radiation-transfer simulations, based on
the MC simulations of DSA by \cite{SB12} and the electron dynamics and radiation transfer 
modules of \cite{BC02,Boettcher13}.  This has been applied to a specific multi-wavelength flare
of 3C~279, namely Flare C of \cite{Hayashida15}, which was selected because of its approximately 
equal flare amplitudes in the optical, X-rays, and GeV $\gamma$-rays.  The evolving spectra can 
be well modeled by assuming that the shock passage through an active region in the jet affects 
primarily the number of electrons accelerated via DSA.  The model predicts well-correlated variability
in the optical and GeV $\gamma$-ray bands and between mm radio and X-ray bands. The mm radio and
X-ray band variability also is well correlated with the optical and $\gamma$-rays, but lags behind
those variations by $\sim 8$ -- 10 hours. Spectral variability in the optical and $\gamma$-ray
bands is expected to be weak, but significant counter-clockwise spectral hysteresis is expected in the 
mm radio and X-ray bands.

\vspace{5pt}
\noindent {\bf Acknowledgements}:
The work of M. B\"ottcher is supported through the South African Research Chairs Initiative
(grant no. 64789) of the Department of Science and Technology and the National Research
Foundation\footnote{Any opinion, finding and conclusion or recommendation expressed in this
material is that of the authors and the NRF does not accept any liability in this regard.}
of South Africa. 
M. Baring is grateful to NASA for support under the {\it Fermi} Guest Investigator Program
through grant 80NSSC18K1711.


\begin{thebibliography}{99}

\bibitem{Abdo11}Abdo, A. A., et al., 2011, ApJ, 727, 129

\bibitem{Ackermann12}Ackermann, M., et al., 2012, ApJ, 751, 159

\bibitem{Baring17}Baring, M. G., B\"ottcher, M., \& Summerlin, E. J.,
2017, MNRAS, 464, 4875

\bibitem{Boettcher97}B\"ottcher, M., Mause, H., \& Schlickeiser, R., 1997, A\&A, 324, 395

\bibitem{Boettcher03}B\"ottcher, M., et al., 2003, ApJ, 596, 847

\bibitem{BC02}B\"ottcher, M., \& Chiang, J., 2002, ApJ, 581, 127

\bibitem{BD10}B\"ottcher, M., \& Dermer, C. D., 2010, ApJ, 711, 445

\bibitem{Boettcher13}B\"ottcher, M., Reimer, A., Sweeney, K., \&
Prakash, A., 2013, ApJ, 768, 54

\bibitem{BB17}B\"ottcher, M., Baring, M. G., \& Summerlin, E. J., 2017, in Proc. of ``HEASA 2016''', PoS, 275, 74

\bibitem{Chen11}Chen, X., Fossati, G., Liang, E. P., \& B\"ottcher, M., 2011,
MNRAS, 416, 2368

\bibitem{Chen12}Chen, X., Fossati, G., B\"ottcher, M., \& Liang, E. P., 2012,
MNRAS, 424, 789

\bibitem{EK88}Edelson, R. A., \& Krolik, J. H., 1988, ApJ, 333, 646

\bibitem{Ellison90}Ellison, D. C., Reynolds, S. P., \& Jones, F. C., 1990,
ApJ, 360, 702

\bibitem{ED04}Ellison, D. C., \& Double, G. P., 2004, Astropart. Phys., 22, 323

\bibitem{Graff08}Graff, P. B., Georganopoulos, M., Perlman, E. S., \& Kazanas, D.,
2008, ApJ, 689, 68

\bibitem{Hayashida15}Hayashida, M., et al., 2015, ApJ, 807, 79


\bibitem{JB11}Joshi, M., \& B\"ottcher, M., 2011, ApJ, 727, 21

\bibitem{KH89}Kirk, J. G., \& Heavens, A. F., 1989, MNRAS, 239, 995

\bibitem{MG85}Marscher, A. P., \& Gear, W. K., 1985, ApJ, 298, 114

\bibitem{Mimica04}Mimica, P., Aloy, M. A., M\"uller, E., \& Brinkmann, W., 2004,
A\&A, 418, 947

\bibitem{Sikora97}Sikora, M., Madejski, G., Moderski, R., \& Poutanen, J., 1997,
ApJ, 484, 108

\bibitem{Sokolov04}Sokolov, A., Marscher, A. P., \& McHardy, I. M., 2004, ApJ, 
613, 725

\bibitem{SM05}Sokolov, A., \& Marscher, A. P., 2005, ApJ, 625, 52

\bibitem{Spada01}Spada, M., Ghisellini, G., Lazzati, D., \& Celotti, A., 2001,
MNRAS, 325, 1559

\bibitem{SB12}Summerlin, E. J., \& Baring, M. G., 2012, ApJ, 745, 63

\bibitem{Takahashi96}Takahashi, T., et al., 1996, ApJ, 470, L89

\end{thebibliography}
\end{document}